\pdfoutput=1

\documentclass[11pt]{article}

\usepackage{acl}

\usepackage{times}
\usepackage{latexsym}

\usepackage[T1]{fontenc}

\usepackage[utf8]{inputenc}

\usepackage{microtype}

\usepackage{algorithm}
\usepackage{algorithmic}
\usepackage{microtype}
\usepackage{multirow}
\usepackage{amssymb}
\usepackage{amsmath}
\usepackage{booktabs}
\usepackage{rotating}
\title{DeltaNet:Conditional Medical Report Generation for COVID-19 Diagnosis}

\author{Xian Wu\textsuperscript{1},Shuxin Yang\textsuperscript{2,5},Zhaopeng Qiu\textsuperscript{1},Shen Ge\textsuperscript{1},Yangtian Yan\textsuperscript{1}, Xingwang Wu\textsuperscript{3}\\ 
\textbf{Yefeng Zheng\textsuperscript{1}, S. Kevin Zhou\textsuperscript{4}, Li Xiao\textsuperscript{2,5}}\thanks{\ \ Corresponding author}\\
\textsuperscript{1}{Tencent Jarvis Lab}
\textsuperscript{2}{Institute of Computing Technology Chinese Academy of Sciences}\\
\textsuperscript{3}{The First Affiliated Hospital of Anhui Medical University}\\
\textsuperscript{4}{University of Science and Technology of China}\\
\textsuperscript{5}{University of Chinese Academy of Sciences}\\
\footnotesize{\{kevinxwu,shuxinyang,zhaopengqiumshenge,yangtianyan,yefengzheng\}@tencent.com;}\\
\footnotesize{duobi2004@126.com; s.kevin.zhou@gmail.com;andrew.lxiao@gmail.com,xiaoli@ict.ac.cn}
}

\begin{document}
\maketitle
\begin{abstract}
Fast screening and diagnosis are critical in COVID-19 patient treatment. In addition to the gold standard RT-PCR, radiological imaging like X-ray and CT also works as an important means in patient screening and follow-up. However, due to the excessive number of patients, writing reports becomes a heavy burden for radiologists. To reduce the workload of radiologists, we propose DeltaNet to generate medical reports automatically. Different from typical image captioning approaches that generate reports with an encoder and a decoder, DeltaNet applies a conditional generation process. In particular, given a medical image, DeltaNet employs three steps to generate a report: 1) first retrieving related medical reports, i.e., the historical reports from the same or similar patients; 2) then comparing retrieved images and current image to find the differences; 3) finally generating a new report to accommodate identified differences based on the conditional report. We evaluate DeltaNet on a COVID-19 dataset, where DeltaNet outperforms state-of-the-art approaches. Besides COVID-19, the proposed DeltaNet can be applied to other diseases as well. We validate its generalization capabilities on the public IU-Xray and MIMIC-CXR datasets for chest-related diseases. Code is available at \url{https://github.com/LX-doctorAI1/DeltaNet}.
\end{abstract}

\section{Introduction}

Since December 2019, the world has been suffering from a serious health crisis: the outbreak of COVID-19~\cite{covid-situation}. 
Fast screening and diagnosis is critical in COVID-19 patient treatment.  In clinical practice, the Reverse Transcription Polymerase Chain Reaction (RT-PCR) is recognized as the golden standard~\cite{2020Coronavirus}. However, due to high false-negative rate and shortage of equipment~\cite{pcr-pb1,pcr-pb2},  medical imaging like X-ray and Computed Tomography (CT) ~\cite{role-img} also works as an alternative means in COVID-19 diagnosis and treatment which generates more timely results than RT-PCR and helps in evaluating the severity degree of COVID-19. 

Given medical images of COVID-19 patients, radiologists need to write relatively long reports to address the impressions and findings. Considering the large volume of COVID-19 patients and potentially infected population, writing medical reports becomes a heavy burden for radiologists. Furthermore, due to the varied expertise of radiologists, some abnormalities in medical images may be ignored and thus not included in the final reports. To alleviate the heavy workload and aid less experienced radiologists, automatically generating medical reports becomes a critical task.

Due to its importance in clinical practice, automatically generating medical reports has attracted extensive research interests in recent years~\cite{zhou2021review}. Existing works mainly follow the image captioning approaches and employ an encoder-decoder process. In the encoding stage, the visual features are extracted from medical images via a CNN; in the decoding stage, the reports are generated sequentially via an RNN. Such a two-step framework has been proven effective in generating general image captions, such as MS COCO. However, when applied to medical images, it may have following two problems: 1) visual bias: for most cases, the abnormal regions only occupy a small fraction of the entire medical images, thus visual features of normal regions dominate the extracted visual embedding, and abnormal regions are difficult to identify; 2) textual bias: in current medical reports, the majority of transcription focuses on describing the normal regions which distract the model from abnormal regions in training.

To address two problems mentioned above, we propose DeltaNet, which is customized for medical images. Different from encoder-decoder frameworks that generate reports from a single medical image, DeltaNet introduces a retrieve-update process which consists of the following three steps: 1) Retrieval: DeltaNet firstly retrieves conditional medical images and reports from medical records. For patients who've already been examined before, we directly obtain his previous medical images and reports. For patients examined for the first time, we retrieve the medical images and reports with similar visual features from other patients; 2) Comparison: DeltaNet compares the embedding difference between two medical images to capture the visual difference; 3) Based on the identified visual difference and the retrieved reports, DeltaNet conditionally generates the final report for current medical image. 

To prove the effectiveness of the proposed DeltaNet, we collect a dataset of COVID-19 patients. For each patient, we manage to collect all the historical medical images and the corresponding reports that cover the complete treatment process. The experimental results show that the proposed DeltaNet outperforms the baselines, including both the general image captioning approaches and the existing medical report generation works. Besides generating reports, DeltaNet can also highlight the differences between previous and current medical images and the corresponding updated reports, which provides a clear explanation for the generated report.
In addition to COVID-19, the proposed DeltaNet can also be applied to medical report generation of other diseases. We evaluate the generalization ability of DeltaNet on IU-Xray and MIMIC-CXR datasets. DeltaNet consistently outperforms baselines. 

\begin{figure*}
  \centering
  \includegraphics[width=\linewidth]{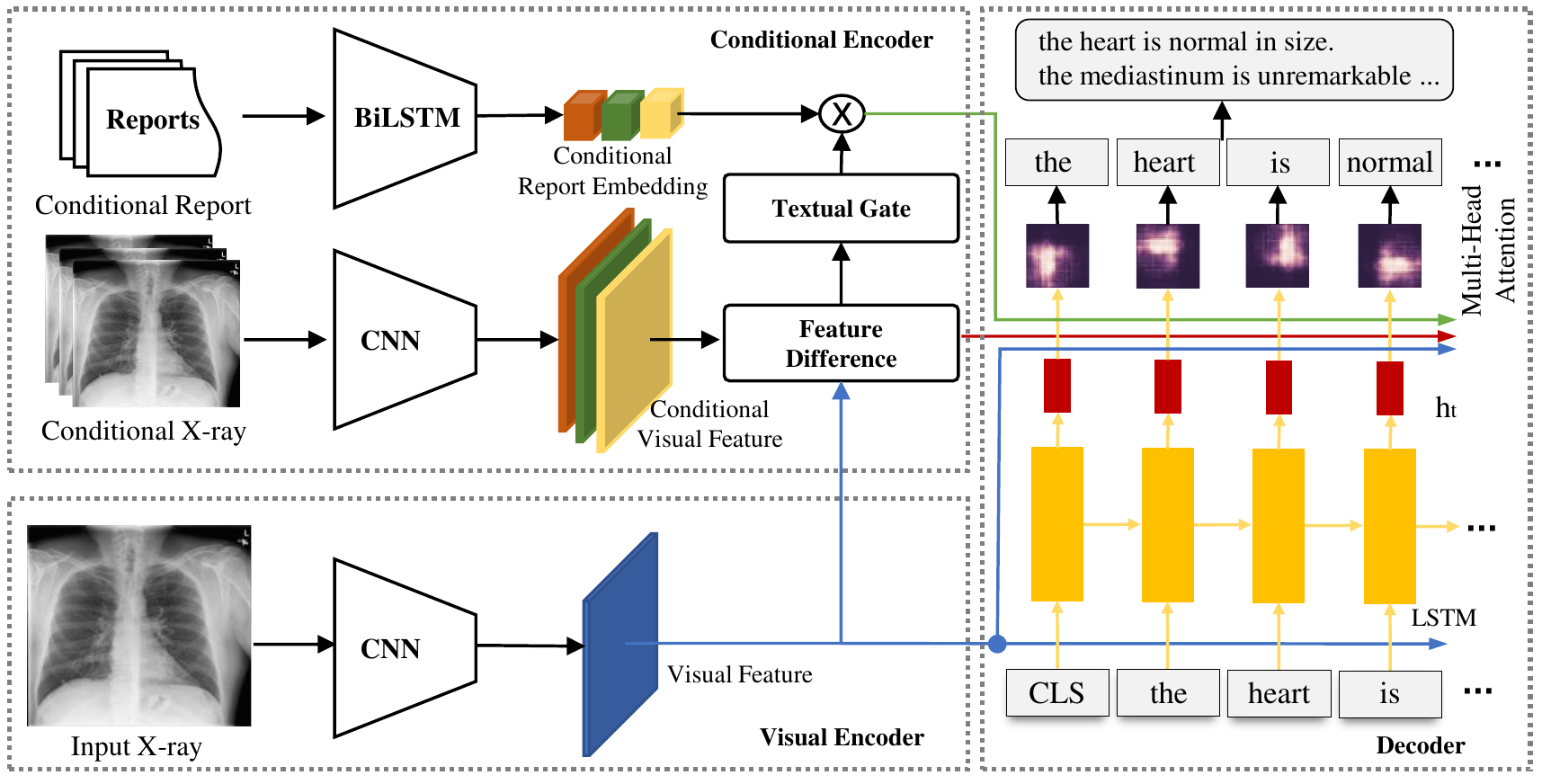}
  \caption{Framework of the proposed DeltaNet model including three major component: visual encoder, conditional encoder and decoder. The visual encoder extracts features from input X-ray image. The conditional encoder extracts features from conditional X-ray image, and embeds the conditional report by a BiLSTM, and acquires the difference between input x-ray and conditional X-ray. The decoder decodes all features, difference and embeddings to generate output report.}
  \label{fig:workflow}
\end{figure*}

\section{Related Works}
\label{sec:related}

\subsection{Image Captioning}
Image captioning aims to provide a short descriptive sentence for a given image, which has recently received extensive research interests \cite{vinyals2015show,xu2015show,anderson2018bottom,Huang2019AoANet}. Typical image captioning models, e.g., \cite{anderson2018bottom,Huang2019AoANet}, adopt the encoder-decoder framework to accomplish the image captioning task, where the encoder extracts the visual representations from the images, and the decoder transforms the acquired visual features to texts. Different from typical image captioning approaches, DeltaNet applies a retrieve-update process, consisting of retrieval, comparison, and generation steps to generate reliable and robust reports.

\subsection{Medical Report Generation}

Medical report generation generally produces longer texts than typical image captions~\cite{zhou2019handbook}.
Employing the typical image caption encoder-decoder framework, \citet{Jing2018Automatic} proposed to use a two-level hierarchical LSTM to deal with long reports, with the top-level handling topic generation and the bottom-level generating texts according to the currently selected topic; \citet{Wang2018TieNet,Xue2019improved,Yuan2019Automatic} used attention mechanism to drive the encoder and the decoder to emphasize on more informative words or visual regions, resulting in improved performance; \citet{Li2018Hybrid,Li2019Knowledge,Zhang2020When,Liu2021Exploring} introduced external information such as template or knowledge graph to guide the generation of medical reports; \citet{Tanveer2020chest} invited four clinicians to manually examine 220,000 reports and build a taxonomy of 11,000 unique terms for developing an automatic labeling algorithm. Then, they built and customized similar reports from a large report database by fine-grained labels as the generated report. However, for diseases not labeled in advance, this method may not fit well. Some other works~\cite{Wang2018TieNet} brought in auxiliary tasks to improve report generation, usually requiring extra expert labeling. Our work uses a new retrieve-update process to effectively generate reports automatically conditioned on historical reports. Besides, our method enhances the model by only using the information from the training dataset, avoiding any expert labeling or external information.

\section{Model}
\label{sec:model}

In this section, we introduce the proposed DeltaNet. Firstly, we formulate the conditional medical report generation problem; secondly, we describe the basic encoder-decoder based generation model; then, we propose DeltaNet, a conditional medical report generation model; finally, we further extend DeltaNet to exploit multiple conditional reports. 

\subsection{Problem Formulation}
\label{sec:formulation}
We use a quadruple $q=\{I, I_c, R, R_c\}$ to refer to an input instance, in which $I$ denotes the input medical image and $R=\{w_1, w_2, \ldots, w_N\}$ denotes the corresponding report to be generated. In this manner, medical report generation can be formulated as estimating the parameters of the conditional probability $P(R|I)$. In this paper, we introduce a conditional image $I_c$ and a conditional report $R_c=\{w_1, w_2, \ldots, w_{N_c}\}$. For patients who have historical medical images, we select their previous medical image and report as $I_c$ and $R_c$ respectively; as for patients without historical reports, we retrieve the most similar medical image $I_c$ and report $R_c$ from the pre-built medical report repository. Then the conditional medical report generation problem can be formulated as $P(R|I,I_c,R_c)$.

\subsection{Basic Model}
\label{sec:basicmodel}

Many existing medical report generation works follow the image captioning approaches and employ the encoder-decoder two-step manner. Typically, a Convolution Neural Network (CNN) is introduced as the encoder, which extracts visual features from the input medical image. In this paper, we extract the output of the last convolution layer of the visual encoder following a linear projection layer as visual features: 
\begin{align}
\label{eqn:visualencoder}
    V & = \text{CNN}(I),
\end{align}
where $V \in \mathbb{R}^{K \times D}$, $K$ denotes the size of feature maps and $D$ denotes the number of feature maps; in this paper, $K$ is set to 49 and $D$ is set to 512.

After acquiring the visual features $V$, the next step is to generate medical reports which we refer as the decoding process.
In the decoding process, we use LSTM to generate medical report sequentially. At each timestamp, we firstly acquire current hidden state:
\begin{align}
\label{eqn:lstm}
    h_t & = \text{LSTM}(w_{t-1}, h_{t-1}),
\end{align}
where $h_0 = \mathbf{0}$ and $h_t \in \mathbb{R}^{1 \times D}$.

After acquiring $h_t$, we attend it to the extracted visual features. In Eq.(\ref{eqn:attend}), we use $h_t$ as query and $V$ as both the key and value:
\begin{align}
\label{eqn:attend}
a_t & = \text{MHA}(h_t, V, V), 
\end{align}
where $\text{MHA}(\cdot,\cdot,\cdot)$ refers to the multi-head attention function proposed in \cite{ashish2017attention}.
The attention embedding $a_t \in \mathbb{R}^{1 \times D}$ can be regarded as attended visual features given the current hidden state.

Finally, we combine the hidden state $h_t$ and the attention embedding $a_t$ to estimate the probability distribution of generating $w_t$. The probability distribution can be formulated as:
\begin{align}
\label{eqn:probability}
P(w_t|w_1,\ldots,w_{t-1},V) & =\sigma([h_t;a_t] W_p),
\end{align}
where $W_p \in \mathbb{R}^{2D \times E}$ is a learnable linear projection and $E$ is vocabulary size.

\subsection{Conditional Generation Model}
\label{sec:con}
In Section \ref{sec:basicmodel}, we adopt the two-step encoder-decoder framework to generate medical reports. Although such a two-step approach has been proven effective in general image captioning, e.g., MSCOCO, due to the unavailability of large-scale labeled data, it is difficult to generate an accurate yet fluent report when applying to the medical report generation problem.

Observing the fact that during a complete treatment process, one patient usually have been examined for multiple times, therefore he may have multiple medical images. For example, the COVID-19 patients may first take a chest X-ray examination for diagnosis and take several more X-ray to track the severity progresses of COVID-19 during the treatment. We review the medical reports from the same patient and find that the consecutive reports share the majority of the content and only differ in the disease progresses. Therefore, a natural thought is to generate the medical report in a conditional manner, not only from the current input medical image but also from the historical images and reports.
For first time patients without previous reports, we can also select reports with similar visual appearances from other patients. For all medical images, we extract their visual features with Eq.(\ref{eqn:conditionimage}). Then given an input image, we retrieve the most similar images according to the cosine distance of visual features.
Let $I_c$ and $R_c$ denote the conditional image and report respectively, then conditional medical report generation problem can be formulated as $P(R|I,I_c,R_c)$. Here we propose DeltaNet as the medical report's conditional generation model.

DeltaNet model first extracts the visual features $V_c$ and textual features $T_c$ from the conditional image and report.
\begin{align}
\label{eqn:conditionimage}
    V_c &= \text{CNN}(I_c), \\
\label{eqn:conditionreport}
    T_c &= \text{BiLSTM}(R_c), 
\end{align}. 

As demonstrated in~\cite{Wu2018}, the subtraction of visual features is an efficient operation to acquire the difference between features. Thus DeltaNet acquires the varied visual features between $V$ and $V_c$ using:
\begin{align}
\label{eqn:variation}
\Delta V_c & = V - V_c, 
\end{align}

In the decoding process, at each timestamp, after generating the hidden state $h_t$ with Eq.(\ref{eqn:lstm}), we attend $h_t$ to both the varied visual features of medical images and the textual features from the conditional report as follows:
\begin{align}
\label{eqn:attenddelta}
s_t & = \text{MHA}(h_t, \Delta V_c, \Delta V_c), \\
\label{eqn:attendreport}
c_t & = \text{MHA}(h_t, T_c, T_c),
\end{align}

Finally, besides the $h_t$ and $a_t$, we utilize $s_t$ and $c_t$ to estimate the probability distribution of generating $w_t$. The probability distribution can be formulated as:
\begin{align}
\label{eqn:conditionprob}
P(w_t|w_1,\ldots,&w_{t-1},V) \\ \nonumber
&=\sigma([h_t;a_t;s_t; c_t]W_p),
\end{align}
where $W_p$ is a learnable linear projection in the shape of $4D \times E$ and $E$ is vocabulary size.

\subsection{Multiple Conditional Generation Model}
\label{sec:multiple}
In the previous subsection, we mainly deal with the cases that the patient only has one conditional report. However, some patients have multiple historical reports. For first time patients without previous reports, we can also retrieve multiple reports according to the visual similarity from other patients. In this subsection, we extend the proposed DeltaNet to fit for multiple conditional reports.

For each input image, we assume that it has $L$ conditional images and reports, we denote them with two sets $\{I_c^{(1)},I_c^{(2)},\ldots,I_c^{(L)}\}$ and $\{R_c^{(1)},R_c^{(2)},\ldots,R_c^{(L)}\}$. DeltaNet firstly extracts the visual features and textual features from conditional images and reports by two encoders:
\begin{align}
\label{eqn:conditionvisual}
    V_c^{(i)} &= \text{CNN}(I_c^{(i)}), \\
\label{eqn:condition}
    T_c^{(i)} &= \text{BiLSTM}(R_c^{(i)}),
\end{align}

Then DeltaNet acquires varied visual features:
\begin{align}
\label{eqn:multivaried}
\Delta V_c^{(i)} & = V - V_c^{(i)}.
\end{align}

As shown in Figure \ref{fig:workflow}, for each textual feature $T_c^{(i)}$ extracted from the conditional report, we add a textual gate to control its contribution since the contribution of each report is different. Here we introduce a gate weight $g_i$ as follows:
\begin{align}
\label{eqn:gate}
    g^{(i)} &= \sigma (W_vV + W_cV_c^{(i)} + b_i),
\end{align}
where $W_v$ and $W_c$ are learnable linear projections. The $g^{(i)}$ takes visual features of both input image and conditional image into consideration which is used to re-weight the textual features:
\begin{align}
\label{eqn:weightedtextualfeature}
    \hat{T}_c^{(i)} &= g^{(i)} * T_c^{(i)}.
\end{align}
DeltaNet further concatenate varied visual features and weighted textual features in a row-wise manner: 
\begin{align}
    \Delta V_c &= [\Delta V_c^{(1)}; \ldots; \Delta V_c^{(L)}], \\
   T_c &= [\hat{T}_c^{(1)}; \ldots; \hat{T}_c^{(L)}],
\end{align}
where $\Delta V_c \in \mathbb{R}^{KL\times D}$ and $T_c \in \mathbb{R}^{N_cL\times D}$. 

At each timestamp, DeltaNet use multi-head attention to generate the subtractive visual attend feature $s_t$ and the context embedding $c_t$.
\begin{align}
\label{eqn:multiattendvisual}
    s_t & = \text{MHA}(h_t, \Delta V_c, \Delta V_c), \\
\label{eqn:multiattendtext}
    c_t & = \text{MHA}(h_t, T_c, T_c),
\end{align}
where both $s_t$ and $c_t$ are in the shape of $1 \times D$.

Then in each time step, the word $w_t$ is generated according to the probability in Eq.(\ref{eqn:multigeneration}).
\begin{align}
\label{eqn:multigeneration}
P(w_t|w_1,\ldots, &w_{t-1},V) \\ \nonumber
&=\sigma([h_t;a_t;s_t;c_t]W_p). 
\end{align}

\section{Experiments}
\label{sec:exp}
In this section, we evaluate the proposed DeltaNet from three perspectives: (1) Whether incorporating conditional reports can bring in performance gain in medical report generation; (2) Whether increasing the number of conditional reports can further improve the performance; (3) Which will perform better, the conditional reports that are acquired from the historical reports of patients themselves or retrieved from other patients according to visual similarity.  

\subsection{Implementation Details}
For first time patients without historical reports, we retrieve conditional reports via the embedding similarity of their corresponding images. The embeddings are acquired from an encoder pre-trained on the ChestX-Ray14 dataset~\cite{chestxray-14}. Since the lengths of conditional reports are different, we use zero-padding for each conditional case to pad each report to a fixed length. We adopt ResNet-152 as our visual encoder for both current input image and conditional images, a two-layer bidirectional LSTM as conditional report encoder, and a single-layer LSTM as medical report decoder. The dimension of the visual feature and hidden states are set to 512. We adopt the Adam optimizer with an initial learning rate of 5e-4 and a mini-batch size of 32. We train the model with cross entropy loss for 100 epochs and early stop strategy is adopted.
\begin{table*}[htbp]
\scriptsize
  \centering
  \caption{The performance of baselines and the proposed DeltaNet on IU-Xray and MIMIC dataset. The conditional reports are extracted from reports of other patients according to visual similarity. For the baseline methods, we cite their performance reported in their papers and reported in \protect\cite{Li2018Hybrid} and \protect\cite{Jing2018Automatic}. The {\em Basic} refers to the one introduced in Section \ref{sec:basicmodel}. }
    \begin{tabular}{c|c|cccccc}
    \toprule
    Dataset & Model & BLEU-1 & BLEU-2 & BLEU-3 & BLEU-4 & CIDEr & ROUGE-L \\
    \midrule
    \multirow{10}[4]{*}{IU-Xray} 
          & S\&T~\cite{vinyals2015show}  & 0.216  & 0.124  & 0.087  & 0.066  & 0.294  & 0.306  \\
          & SA\&T~\cite{xu2015show} & 0.399  & 0.251  & 0.168  & 0.118  & 0.302  & 0.323  \\
          & AdaAtt~\cite{lu2017knowing} & 0.220  & 0.127  & 0.089  & 0.068  & 0.295  & 0.308  \\
          & TieNet~\cite{Wang2018TieNet} & 0.286  & 0.160  & 0.104  & 0.074  & /     & 0.226  \\
          & CoAtt~\cite{Jing2018Automatic} & 0.455  & 0.288  & 0.205  & 0.154  & 0.277  & 0.369  \\
          & R2Gen~\cite{chen2020generating} & 0.470  & 0.304  & 0.219  & 0.165  & /     & 0.371  \\
          & PPKED~\citep{Liu2021Exploring}	& 0.483 	& 0.315 	& 0.224 	& 0.168 	& 0.351 	& 0.376 \\

\cmidrule{2-8}          & Basic & 0.417  & 0.264  & 0.184  & 0.138  & 0.467  & 0.343  \\
          & DeltaNet-1C & 0.470  & 0.307  & 0.224  & 0.175  & \textbf{0.853 } & 0.369  \\
          & DeltaNet-3C & \textbf{0.485 } & \textbf{0.324 } & \textbf{0.238 } & \textbf{0.184 } & 0.802  & \textbf{0.379 } \\
    \midrule
    \multirow{8}[4]{*}{MIMIC} 
          & S\&T~\cite{vinyals2015show}  & 0.256  & 0.157  & 0.102  & 0.070  & 0.063  & 0.249  \\
          & SA\&T~\cite{xu2015show} & 0.304  & 0.177  & 0.112  & 0.077  & 0.083  & 0.249  \\
          & AdaAtt~\cite{lu2017knowing} & 0.311  & 0.178  & 0.111  & 0.075  & 0.084  & 0.246  \\
          & BU\&TD~\cite{anderson2018bottom} & 0.280  & 0.169  & 0.108  & 0.074  & 0.073  & 0.250  \\
          & R2Gen~\cite{chen2020generating} & 0.353 & 0.218 & 0.145 & 0.103 & /     & 0.277 \\
          & PPKED~\citep{Liu2021Exploring}   & 0.360 	& 0.224 	& 0.149 	& 0.106 	& /	& \textbf{0.284} \\

\cmidrule{2-8}          & Basic & 0.335  & 0.206  & 0.138  & 0.100  & 0.156  & 0.263  \\
          & DeltaNet-1C & 0.355  & 0.221  & 0.152  & 0.113  & 0.220  & 0.279 \\
          & DeltaNet-3C & \textbf{0.361 } & \textbf{0.225 } & \textbf{0.154 } & \textbf{0.114 } & \textbf{0.281 } & 0.277  \\
    \bottomrule
    \end{tabular}%
  \label{tab:iu-result}%
\end{table*}%

\subsection{Datasets and Settings}
We introduce three data sets COVID-19, IU-Xray~\cite{Demner-Fushman2016} and MIMIC-CXR~\cite{Johnson2019} to evaluate the proposed DeltaNet. 

The COVID-19 dataset includes 1,261 exams (including both images and reports) from 1,085 patients from mobile field hospitals used in the COVID-19 pandemic. The private information has been removed during the data collection process. Each case includes both the X-ray images and the corresponding reports. The max, median, and mean length of the reports are 180, 69, and 72 words, respectively. Thus the report to be generated are relatively lengthy. The ratio of patients with only one visit is 57.48\%. Among these 1,261 cases, in terms of disease severity, 112 cases are severe, 1,113 cases are general, 30 cases are mild, and 6 cases are not labeled.
More than 166 patients have more than 2 reports; we select the latest one as the prediction target and select the second latest one as the conditional report. We split these 166 patients in the ratio of 7:1:2, that is 116 for training, 16 for validating, and 34 for testing. These three groups have no overlap in patients.

IU-Xray~\cite{Demner-Fushman2016} is a public data set which includes 3,955 radiology reports and 7,470 frontal- and lateral-view chest X-ray images. We follow the split of \cite{chen2020generating} which divides the entire data set into train/validation/test in the ratio 7:1:2.

MIMIC-CXR~\cite{Johnson2019} is the largest dataset for the medical report generation, which contains 377,110 chest X-ray images and 227,827 free-text radiology reports. The dataset contains multi-view images, and we select frontal view images in this work.

For the COVID-19 data set, the conditional report is selected from the previous one of the same patient. As to the IU-Xray data set, we are unable to acquire the historical reports for a patient. Therefore we first acquire the visual feature for each image via a CNN based model pre-trained on the ChestX-Ray14 dataset. Then for each image, we select the conditional reports via cosine similarity of visual feature. For the case in the train/validate/test groups, we select their conditional reports only from the training group. In this case, we avoid the situation that the report to be generated appears as the conditional reports in the training data set. As a result, the label leakage is prevented.

For the MIMIC-CXR data set, we exclude the patients without reports. Then we generate two data sets to evaluate two types of conditional report respectively:1)MIMIC: this data set is the same as original MIMIC-CXR dataset. We follow the standard split provided by~\cite{Johnson2019} to divide the entire data set into train/validate/test and the conditional cases are retrieved from the training set using visual feature similarity,where we employ the ResNet-101 model pre-trained on MIMIC-CXR dataset by torchXrayVision~\cite{Cohen2021TorchXRayVision} as the feature extractor for retrieving similar images; 2) MIMIC-Multi-Visit: we first select the patients with more than three reports as candidates. We choose their latest reports as the prediction target and select three most recent previous reports as the conditional reports. The offical training/validation/testing sets split of MIMIC-CXR is 222,758/1,808/3,269, respectively. From the orignial MIMIC-CXR data set, we select the patients with more than 3 visits, resulting the data set MIMIC-Multi-Visit of 11,978/83/165, respectively. To conduct a fair comparison, during the training of the basic model and existing models, we append the conditional reports into their training set. In this manner, the proposed DeltaNet will not include extra labeled data in training.

We select the popular metrics for natural language generation (NLG) tasks and a specific clinical metric for evaluation. The NLG metrics include BLEU-n~\cite{papineni2002bleu}, CIDEr~\cite{vedantam2015cider}, and ROUGE-L~\cite{lin2004rouge} score. The results are computed by MS-COCO caption evaluation tool~\footnote{https://github.com/tylin/coco-caption} automatically.
The clinical efficacy (CE) metric is proposed by R2gen~\cite{chen2020generating} to quantify the precision, recall, and f1 score of medical terminology described in reference and generated reports. Because the IU-Xray dataset does not provide such labels, we only report CE metric on the MIMIC-CXR dataset. For consistency, we employ the CheXpert~\cite{Irvin2019CheXpert} to extract the labels from generated reports.

\begin{table*}[htbp]
  \centering
  \scriptsize
  \caption{The performance of baselines and the proposed DeltaNet on COVID-19 and MIMIC-Multi-Visit data sets. The conditional reports are extracted from patients' own historical reports. 
  }
    \begin{tabular}{c|c|cccccc}
    \toprule
    Dataset & Model & BLEU-1 & BLEU-2 & BLEU-3 & BLEU-4 & CIDEr & ROUGE-L \\
    \midrule
    \multirow{7}[4]{*}{COVID-19} 
          & S\&T~\cite{vinyals2015show}  & 0.604  & 0.573  & 0.546  & 0.523  & 0.148  & 0.639  \\
          & SA\&T~\cite{xu2015show} & 0.619  & 0.586  & 0.557  & 0.534  & 0.167  & \textbf{0.642}  \\
          & AdaAtt~\cite{lu2017knowing} & 0.617  & 0.583  & 0.553  & 0.529  & 0.150  & 0.625  \\
          & BU\&TD~\cite{anderson2018bottom} & 0.600  & 0.563  & 0.531  & 0.504  & 0.120  & 0.603  \\
          & R2Gen~\cite{chen2020generating} & 0.610  & 0.576  & 0.547  & 0.523  & 0.176  & 0.624  \\
\cmidrule{2-8}          & Basic & 0.614  & 0.566  & 0.530  & 0.503  & 0.256  & 0.610  \\
          & DeltaNet & \textbf{0.664 } & \textbf{0.622 } & \textbf{0.588 } & \textbf{0.561 } & \textbf{0.273 } & 0.635  \\
    \midrule
    \multirow{8}[4]{*}{MIMIC-Multi-Visit} 
          & S\&T~\cite{vinyals2015show}  & 0.300  & 0.176  & 0.111  & 0.074  & 0.066  & 0.246  \\
          & SA\&T~\cite{xu2015show} & 0.284  & 0.170  & 0.114  & 0.080  & 0.046  & 0.234  \\
          & AdaAtt~\cite{lu2017knowing} & 0.310  & 0.180  & 0.114  & 0.075  & 0.068  & 0.240  \\
          & BU\&TD~\cite{anderson2018bottom} & 0.284  & 0.170  & 0.111  & 0.079  & 0.089  & 0.253  \\
          & R2Gen~\cite{chen2020generating} & 0.303  & 0.193  & 0.131  & 0.095  & 0.148  & 0.266  \\
\cmidrule{2-8}          & Basic & 0.325  & 0.196  & 0.129  & 0.091  & 0.102  & 0.248  \\
          & DeltaNet-1C & 0.358  & 0.217  & 0.144  & 0.103  & 0.248  & \textbf{0.270 } \\
          & DeltaNet-3C & \textbf{0.371 } & \textbf{0.228 } & \textbf{0.152 } & \textbf{0.107 } & \textbf{0.301 } & 0.264  \\
    \bottomrule
    \end{tabular}%
  \label{tab:covid-result}%
\end{table*}%

\subsection{Quantitative Results}
\textbf{Baselines}. The Tables display the comparison results between existing works and the proposed DeltaNet. Among the baselines, S\&T~\cite{vinyals2015show}, SA\&T~\cite{xu2015show}, AdaAtt~\cite{lu2017knowing} and BU\&TD~\cite{anderson2018bottom} belong to general image captioning approaches; while CoAtt~\cite{Jing2018Automatic}, TieNet~\cite{Wang2018TieNet}, R2Gen~\cite{chen2020generating} and PPKED~\citep{Liu2021Exploring} focus on the specific medical report generation task. 
Since MIMIC-Multi-Visit only includes patients with more than three images, and COVID-19 is a private data set, we cannot directly cite the results of existing models. Therefore, we re-train existing models on these two data sets. 
For R2Gen, we directly use the implementation~\footnote{https://github.com/cuhksz-nlp/R2Gen} provided by its authors; for CoAtt, we use the third party implementation~\footnote{https://github.com/ZexinYan/Medical-Report-Generation}. For other works, we re-implement them ourselves; The {\em Basis Model} refers to the one described in \textit{Basic Model} section; The {\em DeltaNet} and {\em DeltaNet-1C} refers to the one introduced in \textit{Conditional Generation Model} section; The {\em DeltaNet-3C} refers to the one introduced in \textit{Multiple Conditional Generation Model} section with three conditional reports.

\noindent\textbf{Conditional Reports from Other Patients}.
We evaluate DeltaNet with conditional reports retrieved by the visual feature similarity from other patients. Here we conduct experiments on IU-Xray and MIMIC data sets. Since these two data sets are public data sets and the split of train/validate/test is in the standard manner, for existing works, we directly cite the reported performance. As shown in Table \ref{tab:iu-result}, {\em DeltaNet} consistently outperforms the {\em Basic Model}. This demonstrates that besides using the historical reports of the same patient as the conditional report, selecting similar reports as conditional reports can also boost the performance. As shown in Table \ref{tab:iu-result} and Table \ref{tab:ce}, for the existing methods, the proposed DeltaNet with three conditional reports outperforms almost all baselines on both NLG and clinical efficiency metrics. As to the number of conditional reports, we evaluate DeltaNet with 1 and 3 conditional reports. As shown in both Table \ref{tab:covid-result} and Table \ref{tab:iu-result}, the inclusion of more conditional reports generally increases the performance.

\noindent\textbf{Conditional Reports from the Patients Themselves}.
As shown in Table \ref{tab:covid-result}, {\em DeltaNet} significantly outperforms the {\em Basic Model} which proves the effectiveness of introducing conditional reports in medical report generation. Furthermore, the proposed {\em DeltaNet} outperforms state-of-the-art approaches almost all the metrics.

\begin{table}[htbp]
  \centering
  \scriptsize
  \caption{The performance of clinical efficiency.}
    \begin{tabular}{c|c|ccc}
    \toprule
    Dataset & Model & Precision & Recall & F1 \\
    \midrule
    \multirow{7}[4]{*}{MIMIC} 
          & S\&T  & 0.084  & 0.066  & 0.072  \\
          & SA\&T & 0.181  & 0.134  & 0.144  \\
          & AdaAtt & 0.265  & 0.178  & 0.197  \\
          & BU\&TD & 0.166  & 0.121  & 0.133  \\
          & R2Gen & 0.333  & 0.273  & 0.276  \\
\cmidrule{2-5}          & DeltaNet-1C & 0.460  & 0.353  & 0.376  \\
          & DeltaNet-3C & \textbf{0.470 } & \textbf{0.399 } & \textbf{0.406 } \\
    \bottomrule
    \end{tabular}%
  \label{tab:ce}%
\end{table}%

\subsection{Ablation Study}
\textbf{Self vs. Other Patients}.
Here we conduct an experiment to compare the performance of two types of conditional reports. We use MIMIC-CXR dataset and select the patients with $\geq$2 reports. For each patient, we select two most recent reports, that is 49,180 images from 24,590 patients. The method {\em Self} denotes that the conditional report is acquired from the same patient; the method {\em Others} denotes that the conditional report is retrieved via visual feature similarity. To be fair, the conditional reports of the {\em Self} methods are added into the training set of the {\em Others} method. As shown in Table \ref{tab:compare}, the BLEU-4 is 0.114 vs. 0.101. The conditional report from the same patient outperforms the one from other patients.
\begin{table}[htbp]
\scriptsize
  \centering
  \caption{The performance of two types of conditional reports.}
    \begin{tabular}{c|cccc}
    \toprule
    Conditional Reports & BLEU-3  & BLEU-4  & CIDE-r  & ROUGE-L \\
    \midrule
    Others & 0.140  & 0.101  & 0.179  & 0.278  \\
    \midrule
    Self & \textbf{0.158 } & \textbf{0.114 } & \textbf{0.231 } & \textbf{0.289 } \\
    \bottomrule
    \end{tabular}%
  \label{tab:compare}%
\end{table}%

\begin{figure*}
  \centering
  \includegraphics[width=0.99\linewidth]{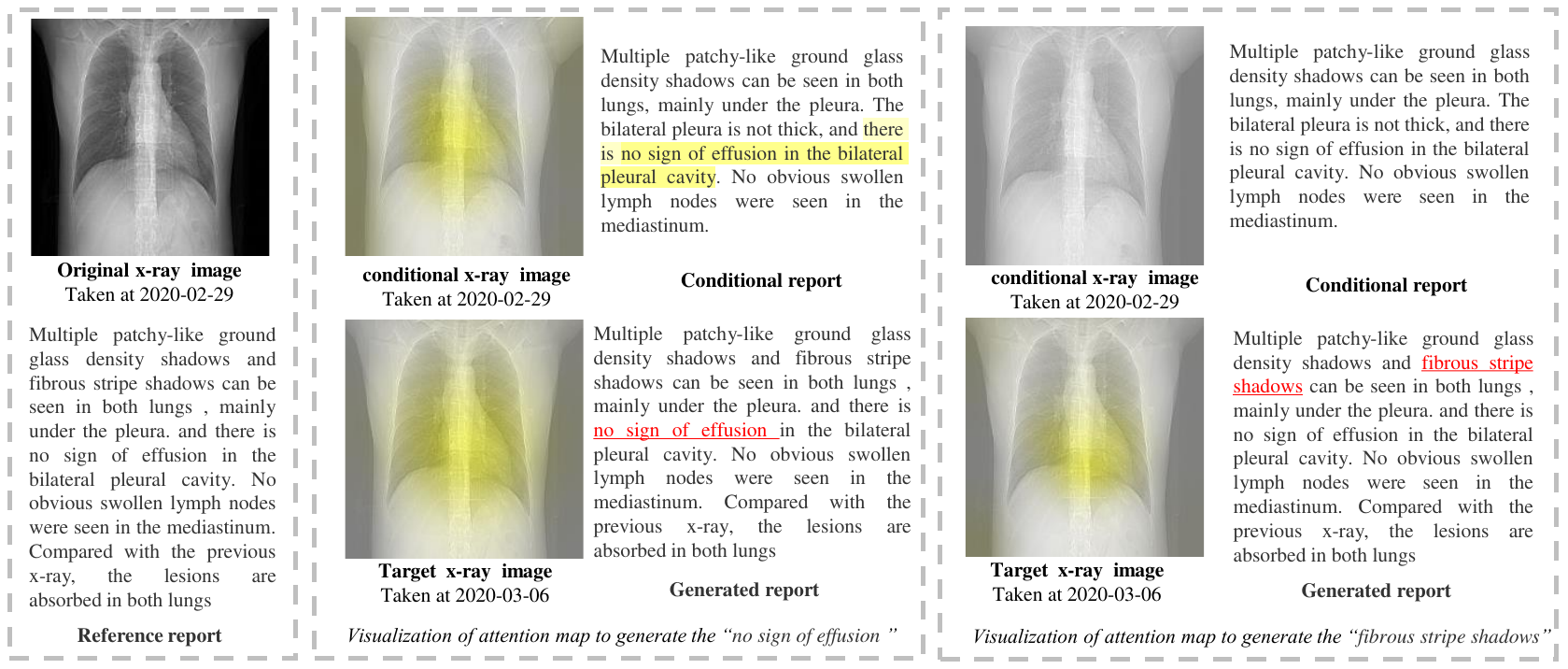}
  \caption{Visualizations of attention map on conditional/target image and report when DeltaNet generates the highlight words. 
  }
  \label{fig:heatmap}
\end{figure*}

\noindent\textbf{Effectiveness of Conditional Image and Report}.
Compared with the basic model, our model utilizes two additional features $I_c$ and $R_c$. To analyze the contribution of each feature, we design two models which separately concatenate $I_c$ or $R_c$ with visual features. As shown in first two rows of Table~\ref{tab:subtraction}, the model with conditional reports outperforms the model with conditional images, which shows that the conditional reports are more effective than conditional images.
To analyze the effectiveness of image feature subtraction in Eq.(\ref{eqn:variation}), we design a model which directly concatenates both with visual features. As shown in ``$I_c + R_c$'' and ours rows of Table~\ref{tab:subtraction}, the model directly concatenates with both features is lower than DeltaNet, which shows the benefit by combining current input and conditional input by feature subtraction and cross attention in DeltaNet.
\begin{table}[htbp]
  \centering
  \scriptsize
  \caption{The contribution of conditional image and report. $I_c$ and $R_c$ refer to the features of conditional images and reports. $I_c + R_c$ refers to feature concatenation.}
    \begin{tabular}{c|c|ccc}
    \toprule
    Dataset & Model & BLEU-3 & BLEU-4 & ROUGE-L \\
    \midrule
    \multirow{4}[4]{*}{IU-Xray} & $I_c$    & 0.137  & 0.103  & 0.281  \\
          & $R_c$    & 0.176  & 0.134  & 0.310  \\
          & $I_c$+$R_c$  & 0.208  & 0.162  & 0.329  \\
\cmidrule{2-5}          & Ours  & \textbf{0.224 } & \textbf{0.175 } & \textbf{0.369 } \\
    \midrule
    \multirow{4}[4]{*}{MIMIC-Multi-Visit} & $I_c$    & 0.108  & 0.075  & 0.222  \\
          & $R_c$    & 0.126  & 0.090  & 0.238  \\
          & $I_c$+$R_c$  & 0.127  & 0.091  & 0.242  \\
\cmidrule{2-5}          & Ours  & \textbf{0.144 } & \textbf{0.103 } & \textbf{0.270 } \\
    \midrule
    \multirow{4}[4]{*}{COVID-19} & $I_c$    & 0.481  & 0.456  & 0.572  \\
          & $R_c$    & 0.507  & 0.482  & 0.595  \\
          & $I_c$+$R_c$  & 0.531  & 0.508  & 0.619  \\
\cmidrule{2-5}          & Ours  & \textbf{0.588 } & \textbf{0.561 } & \textbf{0.635 } \\
    \bottomrule
    \end{tabular}%
  \label{tab:subtraction}%
\end{table}%

\subsection{Case Study}
\label{sec:casestudy}

In this section, we demonstrate the effectiveness of DeltaNet with a case study of a COVID-19 patient. Here we select a patient who has two consecutive X-ray examinations during COVID treatment. This patient is in mild COVID-19 severity. We use the report taken on Feb 29, 2020 as the conditional report to generate the report taken on March 6, 2020.

On the left part of Figure \ref{fig:heatmap}, we visualize the attention heat map to generate the phrase ``no sign of effusion''. The finding of ``no sign of effusion'' remains the same in both conditional and current reports. Therefore, generating this phrase attends to both current and conditional medical images. It also attends to the corresponding phrase in the conditional reports. As a result, the combination of the conditional report and both images enable the correct generation of this phrase.

On the right of Figure \ref{fig:heatmap}, we visualize the attention heat map to generate the phrase ``fibrous stripe shadows''. Since this is a new finding which does not exist in a conditional report, it only attends to the current medical image. The proposed DeltaNet can learn the difference between conditional and current images and correspondingly generate this new finding in the target report.

\section{Conclusion and Future Works}
\label{sec:conclusion}
In this paper, we targeted to automate the medical report generation. Different from typical encoder-decoder framework, we proposed a conditional generation model DeltaNet. For patients with historical reports, we combined the input image with the historical report to generate reports. For first time patients without historical reports, we retrieved visual similar reports from other patients as conditional reports. We proved the advantage of the proposed DeltaNet over state-of-the-art approaches on the IU-Xray, MIMIC-CXR and COVID-19 datasets.

In this paper, for new patients without historical reports, we retrieve the conditional reports only via visual similarity. However, more features like age, gender and diseases could be useful in selecting conditional reports.  In addition, we may attempt to combine historical reports from the same patient and simliar reports from other patients together to futher improve the performance.


\bibliography{anthology,custom}

\begin{thebibliography}{32}
\expandafter\ifx\csname natexlab\endcsname\relax\def\natexlab#1{#1}\fi

\bibitem[{Anderson et~al.(2018)Anderson, He, Buehler, Teney, Johnson, Gould,
  and Zhang}]{anderson2018bottom}
Peter Anderson, Xiaodong He, Chris Buehler, Damien Teney, Mark Johnson, Stephen
  Gould, and Lei Zhang. 2018.
\newblock Bottom-up and top-down attention for image captioning and {VQA}.
\newblock In \emph{{CVPR}}.

\bibitem[{Chen et~al.(2020)Chen, Song, Chang, and Wan}]{chen2020generating}
Zhihong Chen, Yan Song, Tsung-Hui Chang, and Xiang Wan. 2020.
\newblock {Generating Radiology Reports via Memory-driven Transformer}.
\newblock In \emph{EMNLP}.

\bibitem[{Cohen et~al.(2021)Cohen, Viviano, Bertin, Morrison, Torabian,
  Guarrera, Lungren, Chaudhari, Brooks, Hashir, and
  Bertrand}]{Cohen2021TorchXRayVision}
Joseph~Paul Cohen, Joseph~D. Viviano, Paul Bertin, Paul Morrison, Parsa
  Torabian, Matteo Guarrera, Matthew~P Lungren, Akshay Chaudhari, Rupert
  Brooks, Mohammad Hashir, and Hadrien Bertrand. 2021.
\newblock Torchxrayvision: A library of chest x-ray datasets and models.

\bibitem[{Demner-Fushman et~al.(2016)Demner-Fushman, Kohli, Rosenman, Shooshan,
  Rodriguez, Antani, Thoma, and McDonald}]{Demner-Fushman2016}
Dina Demner-Fushman, Marc~D. Kohli, Marc~B. Rosenman, Sonya~E. Shooshan,
  Laritza Rodriguez, Sameer Antani, George~R. Thoma, and Clement~J. McDonald.
  2016.
\newblock \href {https://doi.org/10.1093/jamia/ocv080} {{Preparing a collection
  of radiology examinations for distribution and retrieval}}.
\newblock \emph{Journal of the American Medical Informatics Association},
  23(2):304--310.

\bibitem[{Fang et~al.(2020)Fang, Zhang, Xie, Lin, Ying, Pang, and Ji}]{pcr-pb1}
Yicheng Fang, Huangqi Zhang, Jicheng Xie, Minjie Lin, Lingjun Ying, Peipei
  Pang, and Wenbin Ji. 2020.
\newblock {Sensitivity of Chest CT for COVID-19: Comparison to RT-PCR}.
\newblock \emph{Radiology}, pages 200--432.

\bibitem[{Huang et~al.(2019)Huang, Wang, Chen, and Wei}]{Huang2019AoANet}
Lun Huang, Wenmin Wang, Jie Chen, and Xiaoyong Wei. 2019.
\newblock Attention on attention for image captioning.
\newblock In \emph{{ICCV}}.

\bibitem[{Irvin et~al.(2019)Irvin, Rajpurkar, Ko, Yu, Ciurea{-}Ilcus, Chute,
  Marklund, Haghgoo, Ball, Shpanskaya, Seekins, Mong, Halabi, Sandberg, Jones,
  Larson, Langlotz, Patel, Lungren, and Ng}]{Irvin2019CheXpert}
Jeremy Irvin, Pranav Rajpurkar, Michael Ko, Yifan Yu, Silviana Ciurea{-}Ilcus,
  Chris Chute, Henrik Marklund, Behzad Haghgoo, Robyn~L. Ball, Katie~S.
  Shpanskaya, Jayne Seekins, David~A. Mong, Safwan~S. Halabi, Jesse~K.
  Sandberg, Ricky Jones, David~B. Larson, Curtis~P. Langlotz, Bhavik~N. Patel,
  Matthew~P. Lungren, and Andrew~Y. Ng. 2019.
\newblock Chexpert: {A} large chest radiograph dataset with uncertainty labels
  and expert comparison.
\newblock In \emph{{AAAI}}.

\bibitem[{Jing et~al.(2018)Jing, Xie, and Xing}]{Jing2018Automatic}
Baoyu Jing, Pengtao Xie, and Eric~P. Xing. 2018.
\newblock On the automatic generation of medical imaging reports.
\newblock In \emph{{ACL}}.

\bibitem[{Johnson et~al.(2019)Johnson, Pollard, Berkowitz, Greenbaum, Lungren,
  Deng, Mark, and Horng}]{Johnson2019}
Alistair E.~W. Johnson, Tom~J Pollard, Seth~J Berkowitz, Nathaniel~R Greenbaum,
  Matthew~P Lungren, Chih-ying Deng, Roger~G Mark, and Steven Horng. 2019.
\newblock \href {https://doi.org/10.1038/s41597-019-0322-0} {{MIMIC-CXR, a
  de-identified publicly available database of chest radiographs with free-text
  reports}}.
\newblock \emph{Scientific Data}, 6(1):317.

\bibitem[{Li et~al.(2019)Li, Liang, Hu, and Xing}]{Li2019Knowledge}
Christy~Y. Li, Xiaodan Liang, Zhiting Hu, and Eric~P. Xing. 2019.
\newblock {Knowledge-Driven Encode, Retrieve, Paraphrase for Medical Image
  Report Generation}.
\newblock In \emph{{AAAI}}.

\bibitem[{Li et~al.(2018)Li, Liang, Hu, and Xing}]{Li2018Hybrid}
Yuan Li, Xiaodan Liang, Zhiting Hu, and Eric~P. Xing. 2018.
\newblock {Hybrid Retrieval-Generation Reinforced Agent for Medical Image
  Report Generation}.
\newblock In \emph{{NeurIPS}}.

\bibitem[{Lin(2004)}]{lin2004rouge}
Chin-Yew Lin. 2004.
\newblock Rouge: A package for automatic evaluation of summaries.
\newblock In \emph{Text summarization branches out}, pages 74--81.

\bibitem[{Liu et~al.(2021)Liu, Wu, Ge, Fan, and Zou}]{Liu2021Exploring}
Fenglin Liu, Xian Wu, Shen Ge, Wei Fan, and Yuexian Zou. 2021.
\newblock Exploring and distilling posterior and prior knowledge for radiology
  report generation.
\newblock In \emph{Proceedings of the IEEE/CVF Conference on Computer Vision
  and Pattern Recognition}, pages 13753--13762.

\bibitem[{Lu et~al.(2017)Lu, Xiong, Parikh, and Socher}]{lu2017knowing}
Jiasen Lu, Caiming Xiong, Devi Parikh, and Richard Socher. 2017.
\newblock Knowing when to look: Adaptive attention via a visual sentinel for
  image captioning.
\newblock In \emph{{CVPR}}.

\bibitem[{Ng et~al.(2020)Ng, Lee, Yang, Yang, Li, Wang, Lui, Lo, Leung, Khong
  et~al.}]{pcr-pb2}
Ming-Yen Ng, Elaine~YP Lee, Jin Yang, Fangfang Yang, Xia Li, Hongxia Wang, Macy
  Mei-sze Lui, Christine Shing-Yen Lo, Barry Leung, Pek-Lan Khong, et~al. 2020.
\newblock Imaging profile of the covid-19 infection: radiologic findings and
  literature review.
\newblock \emph{Radiology: Cardiothoracic Imaging}, 2(1):e200034.

\bibitem[{Papineni et~al.(2002)Papineni, Roukos, Ward, and
  Zhu}]{papineni2002bleu}
Kishore Papineni, Salim Roukos, Todd Ward, and Wei-Jing Zhu. 2002.
\newblock Bleu: a method for automatic evaluation of machine translation.
\newblock In \emph{Proceedings of the 40th annual meeting of the Association
  for Computational Linguistics}, pages 311--318.

\bibitem[{Rubin et~al.(2020)Rubin, Haramati, Kanne, Schluger, and
  Wells}]{role-img}
Geoffrey~D. Rubin, Linda~B. Haramati, Jeffrey~P. Kanne, Neil~W. Schluger, and
  Athol~U. Wells. 2020.
\newblock The role of chest imaging in patient management during the covid-19
  pandemic: A multinational consensus statement from the fleischner society.
\newblock \emph{Radiology}, 296(1):201365.

\bibitem[{Syeda{-}Mahmood et~al.(2020)Syeda{-}Mahmood, Wong, Gur, Wu, Jadhav,
  Kashyap, Karargyris, Pillai, Sharma, Syed, Boyko, and
  Moradi}]{Tanveer2020chest}
Tanveer~F. Syeda{-}Mahmood, Ken C.~L. Wong, Yaniv Gur, Joy~T. Wu, Ashutosh
  Jadhav, Satyananda Kashyap, Alexandros Karargyris, Anup Pillai, Arjun Sharma,
  Ali~Bin Syed, Orest~B. Boyko, and Mehdi Moradi. 2020.
\newblock {Chest X-Ray Report Generation Through Fine-Grained Label Learning}.
\newblock In \emph{{MICCAI}}.

\bibitem[{Vaswani et~al.(2017)Vaswani, Shazeer, Parmar, Uszkoreit, Jones,
  Gomez, Kaiser, and Polosukhin}]{ashish2017attention}
Ashish Vaswani, Noam Shazeer, Niki Parmar, Jakob Uszkoreit, Llion Jones,
  Aidan~N. Gomez, Lukasz Kaiser, and Illia Polosukhin. 2017.
\newblock Attention is all you need.
\newblock In \emph{{NIPS}}.

\bibitem[{Vedantam et~al.(2015)Vedantam, Lawrence~Zitnick, and
  Parikh}]{vedantam2015cider}
Ramakrishna Vedantam, C~Lawrence~Zitnick, and Devi Parikh. 2015.
\newblock Cider: Consensus-based image description evaluation.
\newblock In \emph{Proceedings of the IEEE conference on computer vision and
  pattern recognition}, pages 4566--4575.

\bibitem[{Vinyals et~al.(2015)Vinyals, Toshev, Bengio, and
  Erhan}]{vinyals2015show}
Oriol Vinyals, Alexander Toshev, Samy Bengio, and Dumitru Erhan. 2015.
\newblock Show and tell: {A} neural image caption generator.
\newblock In \emph{{CVPR}}.

\bibitem[{Wang et~al.(2020)Wang, Horby, Hayden, and Gao}]{covid-situation}
C.~Wang, P.~W. Horby, F.~G. Hayden, and G.~F. Gao. 2020.
\newblock A novel coronavirus outbreak of global health concern.
\newblock \emph{The Lancet}, 395(10223):470–473.

\bibitem[{Wang et~al.(2017)Wang, Peng, Lu, Lu, Bagheri, and
  Summers}]{chestxray-14}
Xiaosong Wang, Yifan Peng, Le~Lu, Zhiyong Lu, Mohammadhadi Bagheri, and
  Ronald~M. Summers. 2017.
\newblock Chest x-ray8: Hospital-scale chest x-ray database and benchmarks on
  weakly-supervised classification and localization of common thorax diseases.
\newblock In \emph{CVPR}.

\bibitem[{Wang et~al.(2018)Wang, Peng, Lu, Lu, and Summers}]{Wang2018TieNet}
Xiaosong Wang, Yifan Peng, Le~Lu, Zhiyong Lu, and Ronald~M. Summers. 2018.
\newblock Tienet: Text-image embedding network for common thorax disease
  classification and reporting in chest x-rays.
\newblock In \emph{{CVPR}}.

\bibitem[{Wu et~al.(2018)Wu, Liu, Wang, and Dong}]{Wu2018}
Chenfei Wu, Jinlai Liu, Xiaojie Wang, and Xuan Dong. 2018.
\newblock \href {https://doi.org/10.1145/3240508.3240513} {{Object-difference
  attention: A simple relational attention for visual question answering}}.
\newblock In \emph{MM 2018 - Proceedings of the 2018 ACM Multimedia
  Conference}, volume~1, pages 519--527.

\bibitem[{Xu et~al.(2015)Xu, Ba, Kiros, Cho, Courville, Salakhudinov, Zemel,
  and Bengio}]{xu2015show}
Kelvin Xu, Jimmy Ba, Ryan Kiros, Kyunghyun Cho, Aaron Courville, Ruslan
  Salakhudinov, Rich Zemel, and Yoshua Bengio. 2015.
\newblock Show, attend and tell: Neural image caption generation with visual
  attention.
\newblock In \emph{{ICML}}.

\bibitem[{Xue and Huang(2019)}]{Xue2019improved}
Yuan Xue and Xiaolei Huang. 2019.
\newblock Improved disease classification in chest x-rays with transferred
  features from report generation.
\newblock In \emph{{IPMI}}.

\bibitem[{Yuan et~al.(2019)Yuan, Liao, Luo, and Luo}]{Yuan2019Automatic}
Jianbo Yuan, Haofu Liao, Rui Luo, and Jiebo Luo. 2019.
\newblock \href {https://doi.org/10.1007/978-3-030-32226-7_80} {{Automatic
  Radiology Report Generation Based on Multi-view Image Fusion and Medical
  Concept Enrichment}}.
\newblock In \emph{MICCAI 2019}, volume 11769 LNCS, pages 721--729.

\bibitem[{Zhang et~al.(2020)Zhang, Wang, Xu, Yu, Yuille, and
  Xu}]{Zhang2020When}
Yixiao Zhang, Xiaosong Wang, Ziyue Xu, Qihang Yu, Alan Yuille, and Daguang Xu.
  2020.
\newblock \href {https://doi.org/10.1609/aaai.v34i07.6989} {{When Radiology
  Report Generation Meets Knowledge Graph}}.
\newblock \emph{Proceedings of the AAAI Conference on Artificial Intelligence},
  34(07):12910--12917.

\bibitem[{Zhou et~al.(2021)Zhou, Greenspan, Davatzikos, Duncan, Van~Ginneken,
  Madabhushi, Prince, Rueckert, and Summers}]{zhou2021review}
S~Kevin Zhou, Hayit Greenspan, Christos Davatzikos, James~S Duncan, Bram
  Van~Ginneken, Anant Madabhushi, Jerry~L Prince, Daniel Rueckert, and Ronald~M
  Summers. 2021.
\newblock A review of deep learning in medical imaging: Imaging traits,
  technology trends, case studies with progress highlights, and future
  promises.
\newblock \emph{Proceedings of the IEEE}.

\bibitem[{Zhou et~al.(2019)Zhou, Rueckert, and Fichtinger}]{zhou2019handbook}
S~Kevin Zhou, Daniel Rueckert, and Gabor Fichtinger. 2019.
\newblock \emph{Handbook of medical image computing and computer assisted
  intervention}.
\newblock Academic Press.

\bibitem[{Zu et~al.(2020)Zu, Jiang, Xu, Chen, and Zhang}]{2020Coronavirus}
Zi~Yue Zu, Meng~Di Jiang, Peng~Peng Xu, Wen Chen, and Long~Jiang Zhang. 2020.
\newblock Coronavirus disease 2019 (covid-19): A perspective from china.
\newblock \emph{Radiology}, 296(2):200490.

\end{thebibliography}

\newpage

\appendix
\section{Appendix}
\subsection{Ethical Impact}
This work aims to provide efficient and accurate radiology reports to assist radiologists rather than replace radiologists. It benefits the development of human health.
We conduct experiments on COVID-19, IU-Xray, and MIMIC-CXR datasets. All personal information was de-identified. We have removed the information related to the data collection. All necessary permissions have been obtained and the appropriate institutional forms have been archived.

\subsection{Limitation}
In this paper, our method applies a conditional generation process that generates medical reports based on historical medical records. There are still some limitations. First, the generated medical report only contains the diseases in the historical medical records. Second, our method only contains visual and textual information and lacks other examination results (e.g., blood examination and indication) and medical knowledge. We will explore how to involve more features to improve the performance of the proposed method. 

\subsection{Potential Risks}
Given a lot of medical images, it can automatically generate radiology reports. The radiologists only need to make revisions rather than write a new report from scratch. However, inexperienced radiologists may rely on it. Therefore, it is necessary to take additional measures to avoid the abuse of our model.


\label{sec:appendix}


\end{document}